\documentclass[runningheads]{llncs}
\usepackage{graphicx}

\usepackage{tikz}
\usepackage{comment}
\usepackage{amsmath,amssymb} 
\usepackage{color}

\usepackage[accsupp]{axessibility}  

\begin{document}
\pagestyle{headings}
\mainmatter
\def\ECCVSubNumber{100}  

\title{Two-Stage COVID19 Classification Using BERT Features} 



\titlerunning{Abbreviated paper title}
%
\author{Weijun Tan\inst{1}\and
Qi Yao \inst{1}\and 
Jingfeng Liu\inst{1}}
\authorrunning{F. Author et al.}
%
\institute{Jovisin-Deepcam Research Institute, Shenzhen, China
\email{\{sz.twj,sz.yq,sz.ljf\}@jovision.com \\  \{weijun.tan,qi.yao,jingfeng.liu\}@deepcam.com}}
\maketitle

\begin{abstract}

   We present an automatic COVID1-19 diagnosis framework from lung CT-scan slice images using double BERT feature extraction. In the first BERT feature extraction, A 3D-CNN is first used to extract CNN internal feature maps. Instead of using the global average pooling, a late BERT temporal pooing is used to aggregate the temporal information in these feature maps, followed by a classification layer. This 3D-CNN-BERT classification network is first trained on sampled fixed number of slice images from every original CT scan volume. In the second stage, the 3D-CNN-BERT embedding features are extracted on all slice images of every CT scan volume, and these features are averaged into a fixed number of segments. Then another BERT network is used to aggregate these multiple features into a single feature followed by another classification layer. The classification results of both stages are combined to generate final outputs. On the validation dataset, we achieve macro F1 score of 0.9164. 

\end{abstract}

\section{Introduction}

There are a lot of research on automatic diagnosis of COVID-19 since the break-out of this terrible pandemic. 

Among the techniques to diagnosis COVID-19, X-ray and CT-scan images are studied extensively. In this paper, we present an automatic diagnosis framework from chest CT-scan slice images using BERT feature extraction for its extraordinary representing capability with both spatial and temporal attention.  The goal is to classify COVID-19, and non-COVID-19 from a volume of CT-scan slice images of a patient. We use the dataset provided in the ECCV2022 MIA-COV19D challenge \cite{kollias2022mia}, which is the follow-up challenge of the ICCV2021 MIA-COV19D challenge \cite{kollias2021mia}. Since there is no slice annotation in this dataset, 2D CNN classification network on single slice image is not considered. Instead, 3D CNN network based methods are explored.  

A 3D network has been successfully used in many tasks including video understanding (recognition, detection, captioning). A 3D CNN network is much more powerful then 2D CNN feature extraction followed by RNN feature aggregation \cite{kollias2021mia}, \cite{kollias2022mia}. 3D network is also used in COVID-19 diagnosis, where the slice images at different spacing form a 3D series of images. The correlation between slice images is just analogous to the temporal information in videos. 

Transformer is a one of break-through technologies in machine learning in recent years \cite{transformer}. It is first used in language model, then extended to a lot other areas in machine learning. Standard transformer is one directional, and the BERT is a bidirectional extension of it \cite{BERT-original}. It is also first used language model, and later extended to other areas including video action recognition \cite{bert}. The work \cite{WJT-ICCV21} is the first to use a 3D CNN network with BERT (CNN-BERT) for video action recognition \cite{bert} in classification of COVID-19 from CT-scan images.  

We follow the approaches and framework used in \cite{WJT-ICCV21}, including the preprocessing, 3D-CNN-BERT network. Instead of using the MLP for all slice image feature aggregation, we propose to use a second BERT network in the second stage which is more powerful than a simple MLP. The classification results of both stages are combined to generate the final decision for every patient. 

In the first stage, 3D CNN-BERT is used to extract and aggregate a CNN feature for a fixed number of slice images. A classification layer is used on the feature to classify the input to be COVID or non-COVID. In the second stage, we extract the embedding feature vector of all available sets of images for every CT-scan volume. Since the number of images are different, we divide the features into a fixed number of segments (typically 16) using linear interpolation. Then a second BERT is used to aggregate these features into a single feature followed by a few full-connection classification layers. The classification results from both stages are combined to generate the final outputs. 

We evaluate the first stage, the second stage individually and combined. Experiment results show that the combined decision gives best performance. On the validation dataset, we achieve macro F1 score 0.9163.      

\section{Related Work}

Deep learning has been successful in a lot of medical imaging tasks. Some examples are  \cite{kollias2018deep} \cite{kollias2020deep}, \cite{kollias2020transparent}. Since the outbreak of the COVID-19 pandemic, a lot of researches have been done to automatically diagnose on CT scan images or X-ray images. For a latest review, please refer to \cite{Review1}. 

Based on the feature extraction and classification approaches, there are basically three types of methods, the 2D CNN method, 2D+1D method, and 3D CNN method. For the review of this categorization, please refer to \cite{WJT-ICCV21}. 

There are many new developments of 3D CNN for COVID-19 diagnosis recently. We list some of them published in 2022 \cite{SOBAHI2022105335}, \cite{HUANG2022108088}, \cite{RIAHI2022105188}. \cite{BAO2022108499}. Another new method is to use anomaly detection methods to classify COVID-19, including \cite{9187620}, \cite{HASOON2021105045}.

\section{Data Preprocessing and Preparation}

In this section we review the preprocessing and preparation method used in \cite{WJT-ICCV21}, which we use in this work as well. 

\subsection{Slice Image Filtering and Lung Segmentation}

The first task of preprocessing is to select good slice images for training and validation. This is because at the beginning and end of a patient's CT scan images, many images are useless. Some work simply drop a fixed percentage of images at both ends \cite{Tongji}, but we think that method is too coarse. Another motivation for this task is to filter out closed lung images \cite{COVID-CTSet}, because when the lung is closed, the image is also useless. We first use traditional morphological transforms to find the bounding box of the lung, and the mask of the lung contour. We find the max area of lungs out of all slice images of a patient. Then a percentage threshold is set. Slice images whose lung area is less than this thresholds are discarded.  

The second task is to find more accurate lung mask using an UNet. We use an UNet because the morphological lung segmentation is too coarse and miss many important details of the true lung mask, particularly near the edges, and on images with infection lesions. We reuse trained checkpoint in \cite{WJT-ICCV21} to process all the images in the new ECCV22-MIA dataset \cite{kollias2022mia}. After that, a refinement is used to fill the holes and discontinued edges. Shown in Fig. 1 are an example of the preprocessing steps.

\begin{figure}[t]
    \centering
    \includegraphics[scale=0.23]{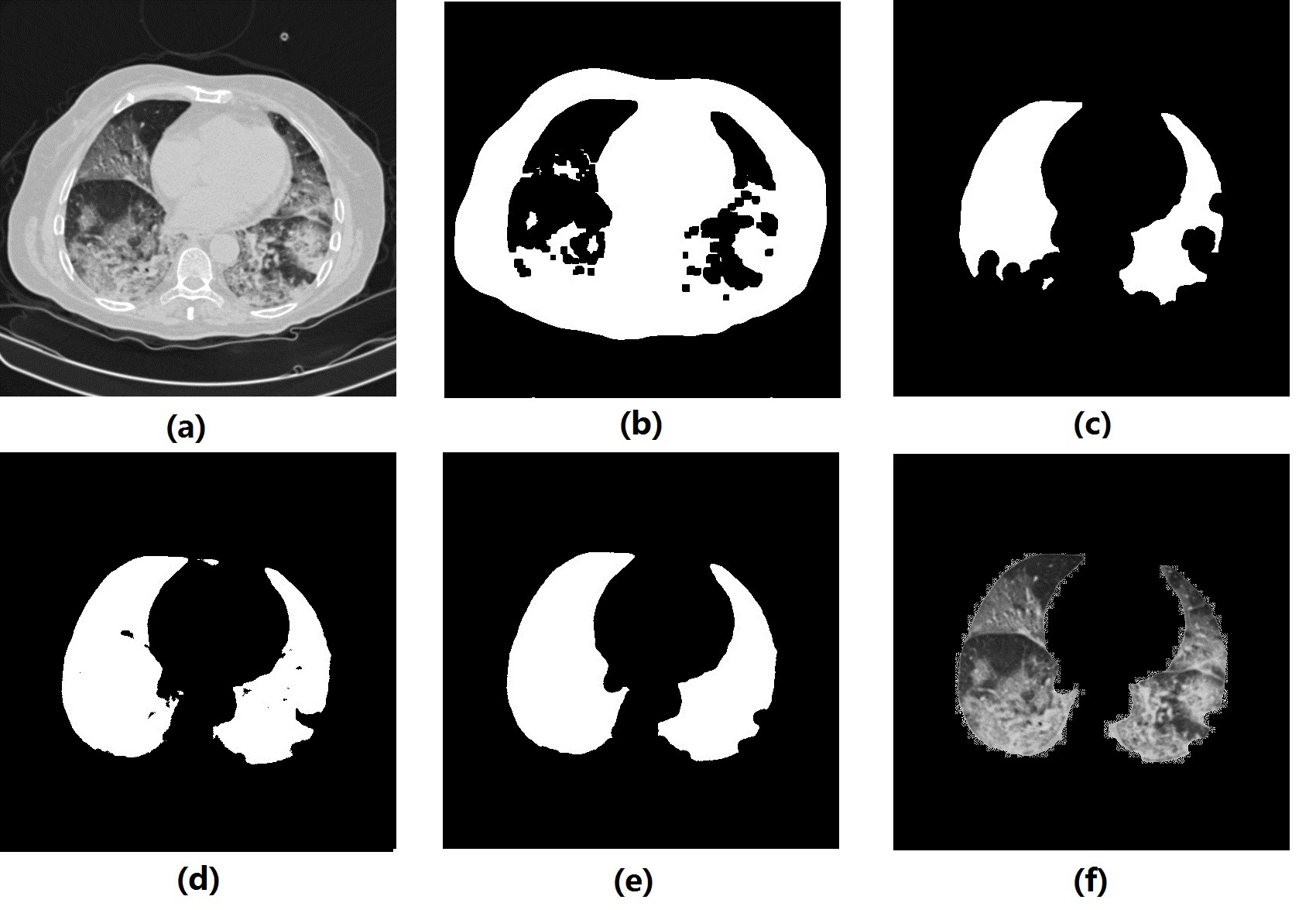}
    \caption{An example of morphological segmentation and UNet segmentation : (a) raw image, (b) after binarization, (c) morphological segmented mask, (d) UNet segmented mask, (e) refined UNet segmented mask, (f) masked lung image. The figure is from \cite{WJT-ICCV21}.}.
    \label{fig1}
\end{figure}

\subsection{Preparing Input for 3D CNN}

The 3D CNN we use requires to use a fixed number of images as input. In our work, we use 32 slice images. However, since the number of available slice images is varying, we need to use resampling strategy to generate the input slice images. There are two cases down-sampling and up-sampling. We use the same resampling method as in \cite{WJT-ICCV21}, which gets the idea from \cite{He2021CovidNet3D}. On the training dataset, random sampling is used, while on the validation and test datasets, a symmetrical and uniform resampling is used. 

In the training, validation and testing time of the first stage 3D-CNN-BERT classification network, only one set of images is selected in every epoch. However, in the second stage BERT classification network, multiple sets of images are selected as much as possible sequentially from the beginning to the end of all slice images of a patient. 

Another big factor of the input is the modality of the input data. An naive idea is to simply use the RGB image. However in \cite{Tongji}, the authors find that adding the segmented mask to the RGB image can help the performance. In \cite{WJT-ICCV21} the authors go further to study the RGB image, the mask and the masked lung image. They find that the combination of these three modalities give the best performance. So in this work we keep using this modality. 

\section{Classification Networks}

In this paper, we explore two stages of classification networks. In the first level, a 3D CNN-BERT network \cite{bert} is used. In the second level, feature vectors of all available set of slices images are generated before the classification layer in the first stage. After equally divided to a given number of segments, a second BERT is used to aggregate these features to a single feature vector for every CT-scan volume. This feature is sent to 3 FC layers to do classification. The diagram is shown in Figure 2. 

\subsection{First Stage 3D CNN-BERT Network}

We reuse the 3D CNN-BERT network in \cite{bert}. This architecture utilizes BERT feature aggregation to replace widely uses global pooling. In this work, we borrow the idea and apply it to COVID-19 classification. In this architecture, the selected 32 slice images are propagated through a 3D CNN architecture and then a BERT feature pooling. In order to perform classification with BERT, an classification token (xcls) is appended at the input and a classification embedding vector (ycls) is generated at the output. This embedding vector is sent the FC layer, producing the classification output. 

In \cite{bert} many different 3D backbone networks are studied, including the R(2+1)D \cite{r2+1d}. On two main video action recognition benchmark datasets, this architecture achieves best performance making new SOTA records. For more detail of this architecture, please refer to \cite{bert}.     
  
In this paper, we use the R(2+1)D backbone \cite{r2+1d} with a Resnet34 \cite{Resnet}. We use 32 slice images as input.  The input image size is set to 224x224, while 112x112 is used in \cite{bert}.  The embedding vector (ycls) is generated and saved for use in the second stage classification network.  

\begin{figure}[t]
    \centering
    \includegraphics[scale=0.375]{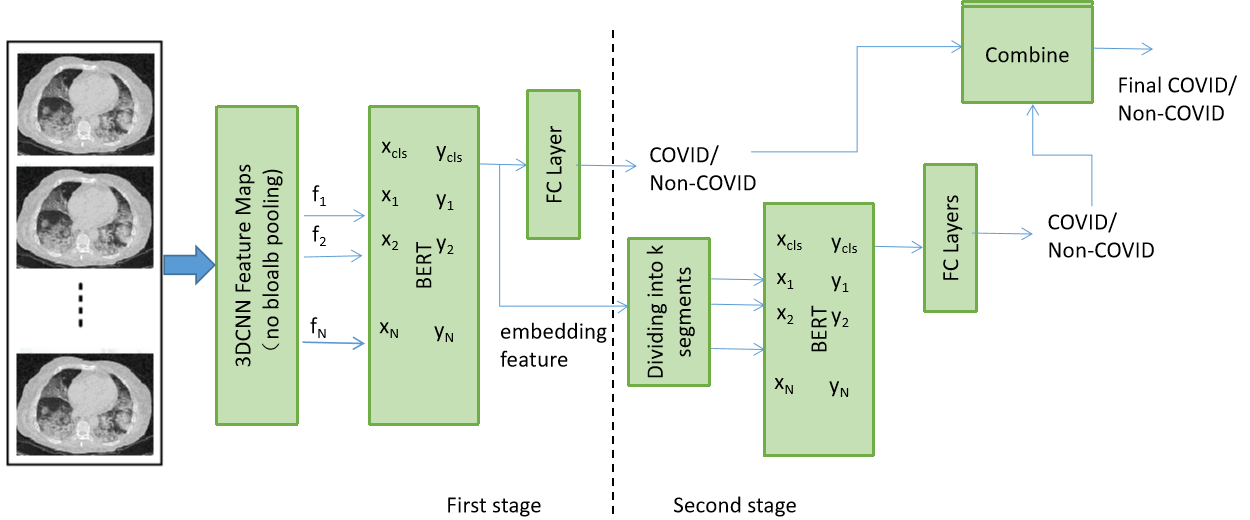}
    \caption{Network architecture of new two-stage classification networks using BERT feature.}
    \label{fig2}
\end{figure}

\subsection{Second Stage BERT Classification Network}

The 3D CNN-BERT network produces a classification result for a single set of input slice images. For CT-volumes where there are more than one set of slice images, we want to process all available slice images in order not to miss some useful information. Therefore, we propose to use a second stage BERT feature aggregation and classification network. 

First, the embedding features are generated for every 32 sequential slice images. If the number of images is less than 32, or the number of images in a sequential set if less than 32, then the same resampling method as in Section 3.2 is used. After that, these features are equally divided into a given number, e.g. 16 segments by taking the mean of features among the available features in that segment. This can be done using simple linear interpolation. Depending the available number of features and number of segments, upsampling or downsampling may be used.      

These given number of segment features are sent to the second BERT network. A classification token is generated inside the BERT, and an output classification vector is sent to the FC layers to do classification. Other outputs are not used. 

\subsection{COVID-19 Severity Classification}

In the second challenge of the ECCV22-MIA, the task is to predict the severity of a COVID-patient. There are four classes: Mild, Moderate, Severe, and Critical. We use the same first stage 3D-CNN-BERT to extract the embedding features. In the second BERT stage, we simply replace the COVID/non-COVID binary classification to the four class severity classification. The standard cross entropy loss instead of the binary cross entropy loss is used.  

\section{Experiment Results}

\subsection{Dataset}

In the ECCV22-MIA dataset \cite{kollias2021mia}, there are 1992 training CT-scan  volumes, 504 validation volumes, and 5544 test volumes. In each volume, there are various number of slice images, ranging from 1 to more than a thousand. There are no slice annotations, so 2D CNN classification is not possible if not using extra datasets. 

Most of the sizes of the images are 512x512. However, there are quite a lot images whose sizes are not so. So we first do a quick check, if the size of an image is not 512x512, it is resized to 512x512 before any other preprocessing is applied.

For the Unet segmentation, we use the annotated dataset we find on Kaggle and from CNBC \cite{CNCB}. In the CNBC annotation, three types of annotations - lung field, Ground-glass opacity, and Consolidation. We merge all three types to the lung field.  

No other datasets are used in training or validation of the 3D CNN-BERT network or the MLP network.  

\subsection{Implementation details}

Both stages of the BERT networks are implemented in one Pytorch framework. In the first stage, the input image size is set to 224x224. We choose to use 32 slice images as input and use the R(2+1)D backbone. In the second stage, the input feature size is 512. Features are equally divided to 4, 8 or 16 segments to find the best performance. We use the Adam optimizer with a initial learning rate 1E-5. A reduced learning rate on plateau with a factor 0.1 and 5 epoch patience is used. The training runs at most 200 epochs with early stopping. The validation accuracy is used to select the best model.   

\textbf{Input Data Modality}: In \cite{WJT-ICCV21}, they study the modality of the slice images. They compare the RGB image only, RGB image and mask \cite{Tongji}, RGM image, mask, and masked image. Their ablation study shows that the last one performs the best. In this work we keep using this modality as input to the first stage 3D-CNN-BERT network. 

\subsection{Detection Challenge Results}

After an ablation study to select parameters, we choose a few top performers to bench mark our algorithm. On the ECCV22-MIA \cite{kollias2021mia} validation dataset, the benchmark results are listed in Table 2. In this table, the F1 scores are presented.   

On the ECCV22-MIA test dataset \cite{kollias2021mia}, the best F1 scores from the first stage 3D CNN-BERT and the second stage BERT classification networks are 88.22\% and 87.4\% respectively.  

\begin{table}
	\begin{center}
		\begin{tabular}{cccc}
		    \hline
			Dataset & 1st stage & 2nd stage & Combined\\
			\hline
			Validation & 0.9073 & 0.9072 & 0.9163  \\
			Test & -  & -  & -\\
			\hline
		\end{tabular}
		\caption{Macro F1 scores of the detection challenge on the ECCV22-MIA datasets.} 
		\label{T3}		
	\end{center}
\end{table}

\subsection{Severity Challenge Results}

On the ECCV22-MIA \cite{kollias2021mia} validation dataset, the severity classification results are listed in Table 3, where the macro F1 scores  presented. Please note that for this task, only the second stage BERT network gives results. We achieve an macro F1 score 0.685. 

\begin{table}
	\begin{center}
		\begin{tabular}{cc}
		    \hline
			Dataset & Macro F1 Score\\
			\hline
			Validation & 0.6855  \\
			Test & - \\
			\hline
		\end{tabular}
		\caption{Macro F1 scores of the severity challenge on the ECCV22-MIA datasets. } 
		\label{T4}		
	\end{center}
\end{table}

\section{Conclusions}

In this paper we present a two stage classification network where in the first stage a 3D CNN-BERT network is used, and in the 2nd stage, a second BERT network is used to aggregate features from all slice images into a single feature for classification. On the validation dataset, our best F1 score is 91.63\%. And on the test dataset, our best F1 score is -.

%
%
\bibliographystyle{splncs04}
\bibliography{egbib}

\begin{thebibliography}{10}
\providecommand{\url}[1]{\texttt{#1}}
\providecommand{\urlprefix}{URL }
\providecommand{\doi}[1]{https://doi.org/#1}

\bibitem{BAO2022108499}
Bao, G., Chen, H., Liu, T., Gong, G., Yin, Y., Wang, L., Wang, X.: Covid-mtl:
  Multitask learning with shift3d and random-weighted loss for covid-19
  diagnosis and severity assessment. Pattern Recognition  \textbf{124},  108499
  (2022)

\bibitem{BERT-original}
Devlin, J., Chang, M.W., Lee, K., Toutanova, K.: Bert: Pre-training of deep
  bidirectional transformers for language understanding. arXiv  (2018)

\bibitem{HASOON2021105045}
Hasoon, J.N., Fadel, A.H., Hameed, R.S., Mostafa, S.A., Khalaf, B.A., Mohammed,
  M.A., Nedoma, J.: Covid-19 anomaly detection and classification method based
  on supervised machine learning of chest x-ray images. Results in Physics
  \textbf{31},  105045 (2021)

\bibitem{Resnet}
He, K., Zhang, X., Ren, S., Sun, J.: Deep residual learning for image
  recognition. CVPR  (2015)

\bibitem{He2021CovidNet3D}
He, X., Wang, S., Chu, X., Shi, S., Tang, J., Liu, X., Yan, C., Zhang, J.,
  Ding, G.: Automated model design and benchmarking of 3d deep learning models
  for covid-19 detection with chest ct scans. Proceedings of the AAAI
  Conference on Artificial Intelligence  (2021)

\bibitem{HUANG2022108088}
Huang, Z., Lei, H., Chen, G., Li, H., Li, C., Gao, W., Chen, Y., Wang, Y., Xu,
  H., Ma, G., Lei, B.: Multi-center sparse learning and decision fusion for
  automatic covid-19 diagnosis. Applied Soft Computing  \textbf{115},  108088
  (2022)

\bibitem{bert}
Kalfaoglu, M.E., Kalkan, S., Alatan, A.A.: Late temporal modeling in 3d cnn
  architectures with bert for action recognition. In: ECCV Workshop. pp.
  731--747. Springer (2020)

\bibitem{9187620}
Karadayi, Y., Aydin, M.N., Öǧrencí, A.S.: Unsupervised anomaly detection in
  multivariate spatio-temporal data using deep learning: Early detection of
  covid-19 outbreak in italy. IEEE Access  \textbf{8},  164155--164177 (2020).
  \doi{10.1109/ACCESS.2020.3022366}

\bibitem{kollias2021mia}
Kollias, D., Arsenos, A., Soukissian, L., Kollias, S.: Mia-cov19d: Covid-19
  detection through 3-d chest ct image analysis. arXiv preprint
  arXiv:2106.07524  (2021)

\bibitem{kollias2022mia}
Kollias, D., Arsenos, A., Soukissian, L., Kollias, S.: Ai-mia: Covid-19
  detection and severity analysis through medical imaging. arXiv preprint
  2206.04732  (2022)

\bibitem{kollias2020deep}
Kollias, D., Bouas, N., Vlaxos, Y., Brillakis, V., Seferis, M., Kollia, I.,
  Sukissian, L., Wingate, J., Kollias, S.: Deep transparent prediction through
  latent representation analysis. arXiv preprint arXiv:2009.07044  (2020)

\bibitem{kollias2018deep}
Kollias, D., Tagaris, A., Stafylopatis, A., Kollias, S., Tagaris, G.: Deep
  neural architectures for prediction in healthcare. Complex \& Intelligent
  Systems  \textbf{4}(2),  119--131 (2018)

\bibitem{kollias2020transparent}
Kollias, D., Vlaxos, Y., Seferis, M., Kollia, I., Sukissian, L., Wingate, J.,
  Kollias, S.D.: Transparent adaptation in deep medical image diagnosis. In:
  TAILOR. pp. 251--267 (2020)

\bibitem{Review1}
Liu, F., Chen, D., Zhou, X., Dai, W., Xu, F.: Let ai perform better next time:
  A systematic review of medical imaging-based automated diagnosis of covid-19:
  2020-2022. Applied Sciences  \textbf{12}(8) (2022)

\bibitem{COVID-CTSet}
Rahimzadeh, M., Attar, A., Sakhaei, S.M.: A fully automated deep learning-based
  network for detecting covid-19 from a new and large lung ct scan dataset.
  medRxiv  (2020)

\bibitem{RIAHI2022105188}
Riahi, A., Elharrouss, O., Al-Maadeed, S.: Bemd-3dcnn-based method for covid-19
  detection. Computers in Biology and Medicine  \textbf{142},  105188 (2022)

\bibitem{MobileNetV2}
Sandler, M., Howard, A., Zhu, M., Zhmoginov, A., Chen, L.: Mobilenetv2:
  Inverted residuals and linear bottlenecks. CVPR  (2018)

\bibitem{SOBAHI2022105335}
Sobahi, N., Sengur, A., Tan, R.S., Acharya, U.R.: Attention-based 3d cnn with
  residual connections for efficient ecg-based covid-19 detection. Computers in
  Biology and Medicine  \textbf{143},  105335 (2022)

\bibitem{WJT-ICCV21}
Tan, W., Liu, J.: A 3d cnn network with bert for automatic covid-19 diagnosis
  from ct-scan images. ICCV Workshops  (2021)

\bibitem{r2+1d}
Tran, D., Wang, H., Torresani, L., Ray, J., Lecun, Y., Paluri, M.: A closer
  look at spatiotemporal convolutions for action recognition. In: CVPR (2018)

\bibitem{transformer}
Vaswani, A., Shazeer, N., Parmar, N., Uszkoreit, J., Jones, L., Gomez, A.N.,
  Kaiser, L., Polosukhin, I.: Attention is all you need (2017)

\bibitem{Tongji}
Wang, X., Deng, X., Fu, Q., Zhou, Q., Feng, J., Ma, H., Liu, W., Zheng, Q.: A
  weakly-supervised framework for covid-19 classification and lesion
  localization from chest ct. IEEE Transactions on Medical Imaging
  \textbf{39},  2615--2625 (August 2020)

\bibitem{SPGC3}
Xue, S., Abhayaratne, C.: Covid-19 diagnostic using 3d deep transfer learning
  for classification of volumetric computerised tomography chest scans. In:
  ICASSP (2021)

\bibitem{CNCB}
Zhang, K., Liu, X., Shen, J., et~al.: Clinically applicable ai system for
  accurate diagnosis, quantitative measurements and prognosis of covid-19
  pneumonia using computed tomography. Cell  (April 2020)

\end{thebibliography}
\end{document}